\documentclass[%
 reprint,
groupedaddress,
frontmatterverbose, 
showpacs,preprintnumbers,
 amsmath,amssymb,
 aps,
 pra,
floatfix
]{revtex4-1}

\usepackage{graphicx,url,graphics}
\usepackage{dcolumn}
\usepackage{bm}
\usepackage{color}
\usepackage{longtable}                                                                            
\usepackage[
   colorlinks,
   linkcolor=blue,
   urlcolor=blue,
   citecolor=blue,
   a4paper=true,
   ]{hyperref}
 
  \def\Nabla{\bigtriangledown}

\begin{document}
 
 
\title{Ultracold, radiative charge transfer in hybrid Yb ion - Rb atom traps}
\author{B. M. McLaughlin}%
\affiliation{
Centre for Theoretical Atomic, Molecular and Optical Physics,
School of Mathematics and Physics, Queen's University Belfast, Belfast BT7 1NN. Northern Ireland, UK.}
\author{H. D. L. Lamb}
\affiliation{
Centre for Theoretical Atomic, Molecular and Optical Physics,
School of Mathematics and Physics, Queen's University Belfast, Belfast BT7 1NN. Northern Ireland, UK.}
\author{I C Lane}
\affiliation{
Innovative Molecular Materials Group, School of Chemistry 
 and Chemical Engineering, Queen's University, Belfast BT7 1NN. Northern Ireland. UK}
\author{J. F. McCann}%
\email{Corresponding author: j.f.mccann@qub.ac.uk}
\affiliation{
Centre for Theoretical Atomic, Molecular and Optical Physics,
School of Mathematics and Physics, Queen's University Belfast, Belfast BT7 1NN. Northern Ireland, UK.}

%

%
%

\date{\today}
 
\begin{abstract}

Ultracold hybrid ion-atom traps offer the possibility of microscopic manipulation of  quantum  coherences in 
the gas using the ion as a probe. However, inelastic processes, particularly  
 charge transfer can be a significant process of ion loss and has been measured experimentally for the 
 Yb$^{+}$ ion immersed in a Rb vapour.
   We use first-principles quantum chemistry codes to obtain the potential energy curves  
   and dipole moments for the lowest-lying energy states of this complex. 
Calculations  for the  radiative decay  processes  cross sections and rate coefficients are presented for the total decay processes; 
Yb$^+$($6s~^2$S) + Rb ($5s~^2$S) $\rightarrow$ Yb($6s^2~^1$S) + Rb$^+$ ($4p^6~^1$S) + h$\nu$ and 
Yb$^{+}$($6s ~ ^2$S) +  Rb($5s~^2$S) $\rightarrow$  YbRb$^{+}$ (X $^1\Sigma^{+}$) + h$\nu$.  
Comparing the semi-classical Langevin approximation with the quantum approach, we 
find it provides  a very good estimate of the background at higher  energies. 
The results demonstrate that radiative decay mechanisms are important 
 over the energy and temperature region considered. In fact,  the Langevin process 
 of ion-atom collisions dominates cold ion-atom collisions. 
 For spin-dependent processes \cite{kohl13} the anisotropic magnetic dipole-dipole interaction
  and the second-order spin-orbit coupling can play important roles, inducing coupling
   between the spin and the orbital motion.  They measured the spin-relaxing collision rate to 
be approximately 5 orders of magnitude higher than the charge-exchange collision rate \cite{kohl13}.
   
Regarding the measured radiative charge transfer collision rate, we find that our calculation 
 is in very good agreement 
with experiment and with previous calculations. Nonetheless, we find no broad resonances 
features that might underly a strong isotope effect.  In conclusion, 
we find, in agreement with previous theory that the isotope anomaly observed in experiment 
remains an open question.

\end{abstract}

 
\maketitle 

 
%
%
%
%
\section{\label{sec:level1}Introduction}

Charge transfer processes in ion-atom collisions  are traditionally
 measured experimentally by their   
cross-sections and rate coefficient as a function of 
energy and temperature. For ambient temperatures,  one can  treat the relative motion of 
the ion and atom as a classical motion and focus on the quantum dynamics 
of the electrons.   However, at ultracold temperatures 
the wave nature of the atomic motion is revealed.
While the electronic and nuclear motion can still be 
adiabatically decoupled, the electronic and nuclear 
motions are strongly correlated. 
Under these conditions, the  chemical pathways and scattering processes 
are highly sensitive to external fields that perturb the electronic structure which in turn 
transfer this effect coherently to the atomic motion. This means that the reaction processes are sensitive 
to external electric and magnetic fields and thus amenable to experimental control, for 
example by Feshbach resonances  ~\cite{foot04,cote98}. 
Such effects are extremely important in controlling coherence 
and correlation, with  applications in  molecular quantum information protocols 
and in hybrid quantum systems such as Coulomb crystals (ion arrays) 
embedded in a quantum degenerate gas,  \cite{juli12,rats12,schm10}.

In this paper, we are concerned with one aspect of  ultracold ion-atom physics: the process of ion loss by charge transfer.
 This is critical in terms  of the ultracold regime as to whether cooling, trapping and degeneracy can be achieved.  It is extremely
important in view of  potential  applications  such as sympathetic  ion cooling and micromotion minimization 
 \cite{hart13}. Furthermore, it allows the study of  the fundamental process of 
ultracold charge transfer ~\cite{fior98,stwa99,Weiner1999}.
 Quantum phenomena  can dominate reaction dynamics at low temperatures. In such cold conditions the scattering 
 process becomes sensitive to the isotopes \cite{lave14} for example when the resonances are sharpened by 
  tunnelling  into long-lived metastable scattering states. 
 
 
Interest has developed in expanding the range of quantum systems that can be trapped and manipulated on the quantum
scale. Hybrid ion-atom systems are of great interest ~\cite{grie09,smit05} since these are inherently strongly-interacting systems 
with a longer-range potential, and inelastic processes can be studied. Recently these systems have been 
explored considering two-body collisions, in which both collision partners are translationally 
cold~\cite{zipk10b}, and on the many-body level~\cite{zipk10a}, where the sympathetic cooling 
of the ion with ultracold atoms was observed. The study of these systems in the quantum regime can be applied to hybrid ion-atom
devices~\cite{idzi07a} and, in addressing fundamental many-body effects of ionic impurities such as
mesoscopic molecule formation~\cite{luki02} and density fluctuations~\cite{gool10}. These devices
offer a unique opportunity to study reactive collisions (ultracold chemistry)  \cite{tacc11} under controlled conditions, for 
example when external electric fields can be applied to modify the reaction rates/cross sections ~\cite{zipk10a}. 
Unlike binary cold collisions between ground state neutral atoms, which are only elastic or inelastic in nature, 
reactive collisions (charge transfer) are a feature of Yb ions immersed in a gas of trapped alkali atoms.
Consequently there has been increased interest in ultracold Yb-ion chemistry in the interactions with alkalis  \cite{zhan10} and 
its resonant charge transfer process \cite{zhan09}. While non-adiabatic effects are strongly 
suppressed as the temperature falls towards zero,  nonetheless 
 the product Yb + Rb$^{+}$ is the thermodynamically favoured species  \cite{lamb12} and thus the 
 loss process can occur by spontaneous emission. It is this process that has received 
 experimental and theoretical attention recently and  we continue to study in detail in this paper.
 
Ultracold neutral atom interactions are characterized by pure $s$-wave scattering mediated at long-range by the
dispersion forces  \cite{hirs54,krems10}. Conversely, a bare ion creates a strong polarization force and hence
the effective cross section is larger with significant contributions from  higher-order partial waves  ~\cite{cote00}. 
Indeed the usual effective range expansion must be  modified by logarithmic terms in the wavenumber expansion~\cite{omal61}.
In the last few years theoretical studies of ultracold ion-atom collisions \cite{zyge14} included the investigation 
of the occurrence of magnetic Feshbach resonances with a view to examining the tunability  of the 
ion-atom interaction focusing on the specific $^{40}{\rm Ca}^{+} - {\rm Na}$ system~\cite{cote03, idzi07b}, 
and calculations of the single-channel scattering properties of the Ba$^+$ ion with the Rb 
neutral atom~\cite{kryc10} which suggest the possibility of sympathetic cooling of the barium 
ion by the buffer gas of ultracold rubidium atoms with a considerable efficiency.
 
In recent experiments ~\cite{zipk10a,zipk10b,dens10}, a single trapped ion of $^{174}$Yb$^+$ in a Paul trap was immersed in a condensate of
neutral $^{87}$Rb atoms confined in a magneto-optical trap.  A study of charge transfer cross sections showed that the 
simple classical Langevin model was inadequate to explain the reaction rates ~\cite{zipk10a}. 
However, very little is known about the microscopic ultracold binary interactions between this ion and the rubidium atom. \cite{juli12,rats12}

The initial experimental study of the quantum coherence of charge transfer ~\cite{zipk10a} 
was analysed using schematic energy curves as no accurate ab initio potentials existed. 
In particular, the potential energy curves and couplings are not known with any  accuracy. 
Thus the experimental study of the quantum coherence of charge transfer ~\cite{zipk10a} 
was based on schematics  of the energy curves.  
This prompted our in depth investigations \cite{lamb12} 
 to map the lowest adiabatic states and the static properties of the molecular ion, 
 in particular the turning points, potential minima, and crossing points of the lowest molecular energies.
In addition to this,  the dissociation energies and molecular constants  provide 
useful spectroscopic data for dynamical investigations \cite{sayf13}. We have 
made a preliminary estimation of the pseudo-potential which approximates the ultracold interaction.
This information is of great importance for modelling ultracold charge transfer, 
and in particular the quantum character of chemical reactivity 
and thus develop insights into ultracold quantum controlled chemistry \cite{krems10}, 
for example when external fields  are applied to influence the reaction rates 
and reaction channels ~\cite{zipk10a}. Of course, the presence of a bare charge in a dilute gas exposes 
many-body physics features such as exciton and polariton dynamics, which are also of great interest.  
It is also of great interest for laser manipulation of the collision to prevent losses 
through charge transfer or create translationally-cold  trapped molecular ions via photoassociation.
Since these processes are light-sensitive, then one can add an extra element of coherent control 
by using a laser to manipulate these processes \cite{hall11,hall13}.
 
In the present study we investigate radiative decay mechanisms, the charge-transfer process
\begin{equation}
\rm Yb^{+} ~(6s ~ ^2S) +  Rb(5s~^2S) \rightarrow Rb^{+}(4p^6~^1S) ~ + Yb (6s^2~ ^1S) + h \nu
\label{proc1}
\end{equation}
and the radiative association process
\begin{equation}
\rm Yb^{+} ~(6s ~ ^2S) +  Rb(5s~^2S) \rightarrow  YbRb^{+} (X ^1\Sigma^{+}) + h \nu
\label{proc2}
\end{equation}
using an optical-potential method.
 
%
%
%
%

\section{Electronic structure calculation}
Following our recent work on this molecular system \cite{lamb12} we extend those computations using a
parallel version of the MOLPRO~\cite{Werner2010} suite of {\it ab initio} quantum chemistry codes 
(release MOLPRO 2010.1) to perform all the molecular structure calculations for this diatomic system (Rb,Yb)$^+$.
Low lying potential energy curves (PEC's) as a function of internuclear distance  out to $R=50$a.u.  are computed and
in the present investigation we extend our earlier work \cite{lamb12} to
 calculate the transition dipole moments between the $^{1,3}\Sigma^{+}$ states. 
 As in our previous work we use an active two-electron model within a
 multi-reference configuration interaction (MRCI) and a full-configuration 
 interaction (FCI) framework to calculate all the potentials.  
Briefly, within the MRCI model,  the  potential energy curves (PEC's) are calculated using effective core 
potentials  (ECP) to replace the non-valence electrons (ECP68MDF for Yb, ECP36SDF for Rb),
 as a basis set for each atom, which allows for  scalar-relativistic effects to be included explicitly. 
 The scalar-relativistic effects are included by adding the corresponding terms 
of the Douglas-Kroll Hamiltonian to the one-electron integrals. 
To model the valence electrons, we use an augmented-correlation-consistent 
polarized valence basis set; aug-cc-pV6Z. We note that the basis set used
yielded values consistent with those of Meyer and Bohn~\cite{meye09} for the neutral YbRb molecule .
To take account of short-range interactions  we employed the 
non-relativistic complete-active-space self consistent field (CASSCF)/MRCI method~\cite{Werner1985,Knowles1985} 
available within the MOLPRO~\cite{Werner2010} {\it ab initio} quantum chemistry suite of codes.  
\begin{figure}
\includegraphics[width=0.525\textwidth]{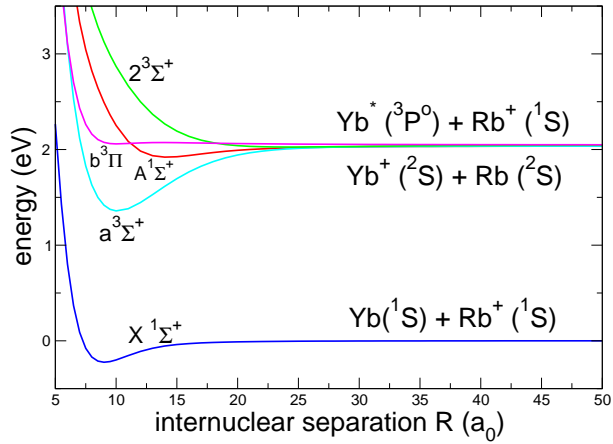}
\caption{Relative electronic energies for the diatomic molecular ion ${\rm YbRb}^+$ as a 
               function of internuclear distance R(a$_0$), (MRCI approximation). The ${\rm X}^{1}\Sigma^{+}$ 
               ground state is the Rb$^{+}$ channel, while the lowest energy ionic ytterbium states, 
                the triplet $a ^{3} \Sigma^{+}$ and singlet $A ^{1}\Sigma^{+}$ pair, are nearly degenerate 
                with the excited charge-transfer channels: $ {\rm Rb}^{+}+{\rm Yb}^{*}$ (see appendix for numerical values).}
\label{fig1}
\end{figure}
\begin{figure}
\includegraphics[width=0.525\textwidth]{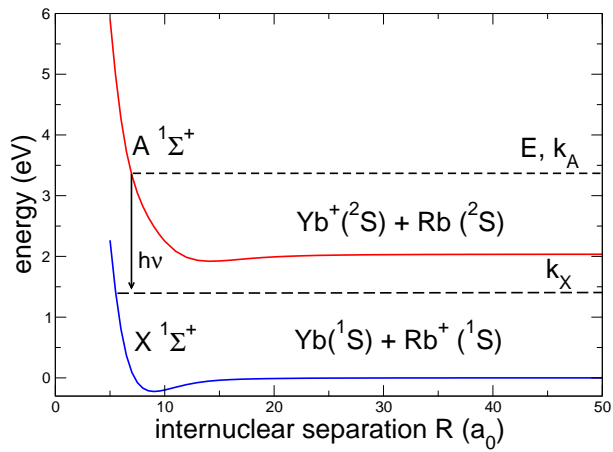}
\caption{${\rm YbRb}^+$  potential energy curves (relative electronic energies) as a function of internuclear 
                 distance R (a$_0$), (MRCI approximation) for the $X ^{1}\Sigma^{+}$  and $A ^{1}\Sigma^{+}$ states.
                The singlet $A ^{1}\Sigma^{+}$  is the entrance channel leading to the radiative charge-transfer 
                channels: $ {\rm Rb}^{+} + {\rm Yb}$, the lowest energy ionic ytterbium states.  
                The ${\rm X}^{1}\Sigma^{+}$  correlates for large $R$ with the Rb$^{+}$ ion. The 
                loss process from state  $A ^{1}\Sigma^{+}$ can be through radiative association into the 
                 bound rovibrational manifold or above the dissociation threshold into the ion-atom charge exchange.}
\label{fig2}
\end{figure}
\begin{figure}
\includegraphics[width=0.525\textwidth]{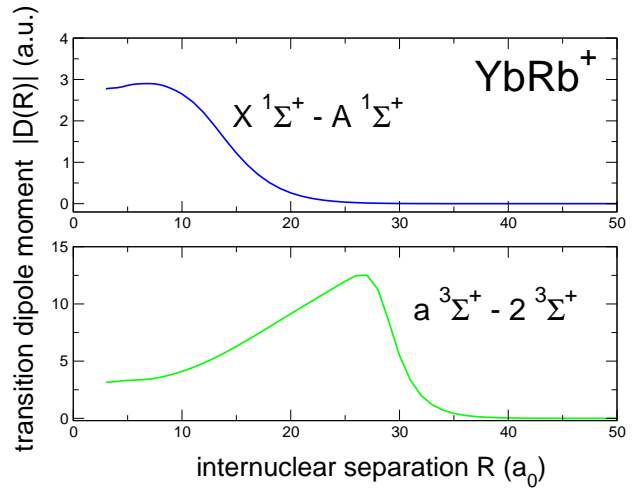}
\caption{Dipole transition moments (absolute value) $\vert D(R) \vert $ for the ${\rm X}^{1}\Sigma^{+} \leftrightarrow
 		{\rm A} ^{1}\Sigma^{+}$ transition and the  ${\rm a}^{3}\Sigma^{+} \leftrightarrow 2 ^{3}\Sigma^{+}$ states
 		as a function of internuclear distance R (a$_0$). The  multi-reference-configuration-interaction (MRCI) 
		approximation within the MOLPRO suite of codes 
		 is used to calculate the transition dipole moments.}
\label{fig3}
\end{figure}
\begin{figure}
\includegraphics[width=0.525\textwidth]{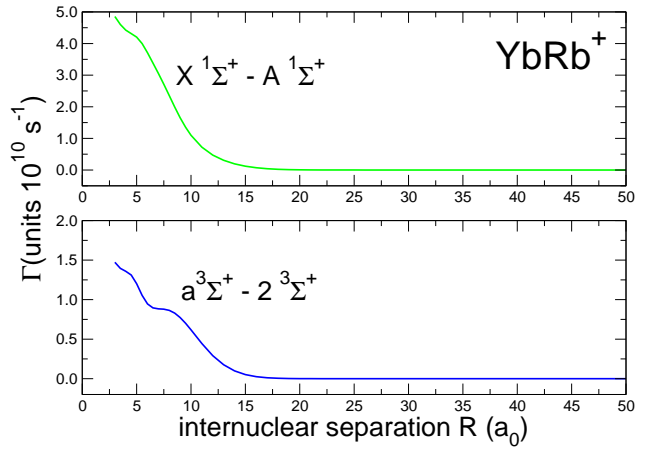}
\caption{Einstein spontaneous emission transition rate $\Gamma (R)$ as a function of internuclear distance R (a$_0$).
		The radiative decay  rate $\Gamma (R)$ (units of 10$^{10}$ s$^{-1}$), according to (\ref{trate}) 
		is shown as a function of internuclear distance R (a$_0$),
		for the ${\rm X}^{1}\Sigma^{+} \rightarrow {\rm A} ^{1}\Sigma^{+}$ 
		states and the ${\rm a}^{3}\Sigma^{+} \rightarrow 2 ^{3}\Sigma^{+}$ transitions.}
\label{fig4}
\end{figure}
Fig. \ref{fig1} shows our adiabatic  potential curves for this system as a function of the internuclear separation $R$.
We note that the present quantum chemistry calculations and those of  Sayfutyarova et al.  \cite{sayf13} use a similar approach.
In summary, both calculations essentially use an effective core-potential 
and a multi-reference CI to cater for electron-correlation in the outer electrons.
At short bond lengths all the results are obtained from the state-averaged CASSCF/MRCI approach. 
Our results are very  similar to those obtained by Sayfutyarova et al \cite{sayf13} who used a total of 22 correlated 
electrons (14 electrons in doubly occupied orbitals). These authors also conducted CCSD(T) calculations 
on the $X^1\Sigma^+$, $a^3\Sigma^+$ and $b^3\Pi$ states over 30 electrons to refine these potentials further. 
The effect of including extra electron correlation leads to a general reduction in the equilibrium bond lengths, 
most significant in the case of the $b^3\Pi$ state, where the combination of the CCSD(T) method and larger 
active electron calculation, predicts a much deeper potential well than found in our present work. However, this well 
nearly halved in size when spin-orbit effects were included. The Yb polarizability determined from the active two 
electron, MRCI and FCI potentials, is 128.5 a.u., (within 8\% of the currently accepted value $139 \pm 7$ \cite{belo12}) 
but smaller than the considerably more expensive CCSD(T) calculations of Sayfutyarova et al.  \cite{sayf13} 
who obtained $\alpha=142.2$.  We note that both calculated values are within the experimental  limits.

In Fig. \ref{fig2} we illustrate the two singlet states involved in the radiative decay processes. 
The radiative charge transfer occurs along  the $A^1\Sigma^+$ state which has a shallow well. 
Comparing our results with the equivalent MRCI potential of Sayfutyarova et al.  \cite{sayf13} 
their results are in good agreement with ours.
 For this well,  our earlier work  \cite{lamb12} found a  dissociation energy was $D_{e}=0.1085$ eV with 
a bond length of  $14.36 a_{0}$, compared to their calculations, where : $D_{e} = 0.1037$eV and  $R_{e}= 13.8139 a_{0}$.
Although our excited state well is slightly deeper, there are significant differences in the $X^1 \Sigma^{+}$  ground-state.   
The dissociation energy of Sayfutyarova et al. \cite{sayf13} is 3496 cm$^{-1}$ (0.4334 eV), almost twice  
 the value of Lamb and co-workers  \cite{lamb12}, who obtained a value of 0.2202 eV. 
 We note also that the equilibrium distance $R_{e}$ also occurs at a shorter bond 
 length of $8.088a_{0}$ \cite{sayf13}, compared to our finding which gave  $R_{e}=9.031 a_{0}$ \cite{lamb12}. 
   
Fig. \ref{fig3}  illustrates the dipole transition moment $D(R)$ (a.u.) 
as a function of internuclear separation $R$ for the singlet and triplet 
$\Sigma^{+}$ states. Results for  the dipole matrix elements  
between the ${\rm X}^{1}\Sigma^{+}$ --  $A ^{1}\Sigma^{+}$ states 
and the ${\rm a}^{3}\Sigma^{+}$ --  $2 ^{3}\Sigma^{+}$ states are illustrated. Comparing these results 
with the work of Sayfutyarova et al. \cite{sayf13} we find very good agreement qualitatively for the $A-X$ moment, 
although it is not possible to compare the triplet-triplet transition moment. Our results 
find a very smooth $A-X$ singlet dipole which leads us to conclude that the sensitivity 
to the dynamics will be due to the wave function envelope. If there were some 
oscillatory behaviour in Fig. \ref{fig3}, then one could anticipate that this might  
be transferred to the radiative coupling. However, in the absence of structures 
in the moment, the resonance behaviour will be primarily potential scattering.

 In Fig. \ref{fig4} the transition rate  calculated using Eq. \ref{trate} is presented as a function of 
the internuclear separation $R$. The decay rate $\Gamma (R)$  decreases exponentially as  $R$ increases due 
to the exponential attenuation in the overlap of the atomic wave functions corresponding to charge transfer. 
Beyond $R$ = 50 a.u., the potential of the  $A ^{1}\Sigma^{+}$ state can be described by the long-range multipole expansion:
\begin{equation}
  V_A(R)=V_{A}(+\infty)- \frac{1}{2}\left[ \frac{\alpha_d}{R^4} +\frac{ C_6}{R^6}+ \frac{C_{8}}{R^8} \right],
\label{mult}
\end{equation}
where  $\alpha_{d}$ is the  dipole polarizability of the neutral atom and
where $C_6$ and $C_8$ are respectively the quadrupole and octupole polarizabilities, which have been 
evaluated in our previous study \cite{lamb12}. In our calculation of the phase shift we integrate into the 
asymptotic regime using the multipole series for the potential.
%
%
%
%

\section{Theoretical Method}

In the simple classical model \cite{Miller70}, the nuclear motion takes place on the incoming 
potential surface, $V_A(R)$. Thus the motion is  angular-momentum 
conserving, time-reversal invariant, and elastic - to a first approximation. 
 Defining the collision energy, in the centre-of-mass frame, as $E$ 
and the reduced mass of the nuclei  as $\mu$,  then we can take the zero 
of potential energy  at infinite separation in the incoming channel: 
$V_{A}(+\infty) =0$. Since angular momentum 
is conserved, then for  an impact parameter $b$, the radial velocity can be 
written as:
\begin{equation}
v_{R}^{2}(R) =  {2 E \over  \mu}  \left(  1 - { V_A(R) \over E} -{ b^2 \over R^2}  \right)    \quad .
\end{equation} 
Thus the classical turning point will be the (largest) solution of the equation:
\begin{equation}
{dR(t) \over dt} = v_{R}(R_c)=0 \quad .
\end{equation} 
The process of spontaneous emission has a rate $\Gamma(R)$ which drives 
the charge transfer process.  Consider a classical trajectory for  a given 
collision energy, $E$, and impact parameter. Then for the decay process, 
letting $t=0$ denotes the classical turning point where $t = \pm \infty$ are 
the end points.  One can write for the probability of emission, for example as explained in  \cite{Miller70}:
\begin{equation}
P(b,E) =  1 -\exp \left( - 2 \int_{R_c}^{+\infty} { \Gamma (R)  \over v_{R}(R)} dR \right)   \qquad .  
\end{equation}
Then to a good approximation, in the case of weak coupling, we have:
\begin{equation}
P(b,E) \approx  2 \int_{R_c}^{+\infty} { \Gamma (R)  \over v_{R}(R)} dR    \qquad .  
\end{equation}
Therefore, the semi-classical cross-section is simply,
\begin{equation}
\sigma_c  (E)  = 2\pi \int_0^{+\infty} b P(E,b) db \quad , 
\end{equation}
which leads to the expression \cite{zyge88}
\begin{equation}
 \sigma_c (E) =  2 \pi \sqrt{\frac{2\mu}{E}} \int_0^{+\infty} b~ db \int_{R_c}^{\infty} \frac{ \Gamma (R) \ dR}{\sqrt{1 - V_A(R)/E - b^2/R^2}}
 \label{semic}
 \end{equation}
 At high energies,  $E \gg V_A$, the integrand in (\ref{semic}) is energy independent  and thus $\sigma (E) \sim (\mu/E)^{1/2}$.
 So for the heavier mass since we are considering the $^{172}$Yb$^{+}$ and  $^{174}$Yb$^{+}$ isotopes  (ignoring 
 any resonant behaviour) then on dynamical grounds the cross-section is slightly higher. 
 It is purely by coincidence  that this  energy dependence  matches the  classical Langevin model \cite{lang05} for reactive collisions. 
For a long-range ion-atom potential we have the polarization potential, $V(R) =-\alpha_{d}/(2R^4)$. 
In the Langevin model,   if the centrifugal barrier can be surmounted then the reaction proceeds with certainty and 
 the cross-section is given by the simple formula:
 \begin{equation}
\sigma_L = \pi \sqrt{ {2 \alpha_d \over E}   }  \qquad ,
\label{lclass}
\end{equation}
which displays the same energy dependence as (\ref{semic}) but based on completely different physics.

Strictly speaking we have three quantum fields: the active electron, the nuclear motion, and the photon. 
One can construct the wavefunctions in the product (adiabatic) representation and then couple 
these through the Hamiltonian (including the vacuum photon states).  However the process 
involves a weak-coupling, the irreversible spontaneous emission leading to charge transfer. 
Thus the collision  of the Yb$^{+}$ ion with the Rb atom leading to loss of the Yb$^{+}$ ion 
can be considered as a second-order perturbation of the elastic lossless collision.   
The modified optical potential will have an imaginary (non-Hermitian) term proportional 
to  the Einstein coefficient. This `width'  depends on the dipole moment 
and frequency of emission and is $R$-dependent.
 The optical potential method, in the context of radiative charge transfer, 
 has been described in detail by Zygelman and Dalgarno\cite{zyge88}.
 We simply present the outline of the main equations and how it is modified for our application.
 
%
%
%
%

 In the adiabatic approximation the dynamics occur on decoupled, centrally-symmetric potential energy curves.  
The temperatures  are so low that all  non-adiabatic radial and rotational coupling are so weak that  the 
vacuum coupling (by photoemission) is the only non-elastic process. 
Radiative charge transfer requires the optical dipole selection rules to 
 be obeyed for transitions to the $^1\Sigma^{+}$ state.  
 Thus only the  $A^1\Sigma^{+}$ state has an allowed radiative charge transfer. 
 
 Using conventional notation,  we use $E$ to denote the collision energy in the  centre-of-mass frame, and 
with $m_i$ and $m_a$ denoting the ion and atom masses, respectively, the reduced mass is defined:
 $\mu=m_i m_a / (m_i + m_a)$. Then the collision wavenumber is denoted by $k=\sqrt{2\mu E}$. 
 Finally,  all potential energies are with respect to  the asymptotic incoming channel:
  $V_A(+\infty)=0$.
 %
 %
 %
 %
 This means that the physics is essentially reduced to a 
 single channel (effective complex radial potential) scattering problem:
 \begin{equation}
 V_A(R) \rightarrow V_A(R)- \textstyle{1 \over 2} i~ \Gamma (R) \quad .
 \end{equation} 

In the  use of the simple optical potential we implicitly, and approximately, take into account both the process 
of radiative association and radiative transfer. That is the lower (exothermic charge exchange) state 
has an (infinite) number of bound (association) rovibrational levels and continuum states. This point is discussed 
in detail in previous applications \cite{zyge14,sayf13} and its validity verified. In other terms,  $\Gamma$, which  
is larger the higher the photon frequency,  is taken as a vertical transition in analogy to  the 
way that the `reflection principle'  is applied \cite{schi93}. This approximation is better the larger the mass 
of the colliding atoms/ions. 
The problem can be summarized mathematically as  \cite{zyge14}  the solution of the 
 Schr\"odinger equation,
\begin{equation}
\left[ -\frac{1}{2\mu} \Nabla^2_{\bf R} + V_A (R) - E \right]  F_A(E;{\bf R}) = \frac{1}{2} i~ \Gamma (R) F_A(E;{\bf R}) 
\end{equation}
where $\Gamma (R)$ is the Einstein spontaneous emission transition rate 
for the decay  $A^1\Sigma^{+} \rightarrow X^1\Sigma^{+}$.
Again using atomic units, we have that:
\begin{equation}
\Gamma (R) = \frac{4 D^2 (R)}{3c^{3}}  \left| V_A(R) - V_X (R) \right | ^3
\label{trate}
\end{equation}
where $c$ is the speed of light, $V_A(R)$ and $V_X (R)$ are the adiabatic potential energies of the upper A $^1\Sigma^+$ and 
ground (lower) X $^1\Sigma^{+}$ states respectively. $D(R)$ is the transition-moment matrix element between the A $^1\Sigma^{+}$ and 
 the X $^1\Sigma^+$ states. For large $R$ values the A $^1\Sigma^{+}$ state separates asymptotically into the 
 atomic states $\rm Yb^{+}(6s~^2S)$ and $\rm Rb(4p^65s ~^2S)$, while the X $^1\Sigma^{+}$ separates into 
 $\rm Yb(6s^2~^1S)$ and $\rm Rb^+(4p^6 ~^1S)$.
Thus, $\Gamma(R)$ is short range  and exponentially damped with increasing $R$ 
  since it requires the electron to transfer from the atom to the ion.  
  As the potential is central, even though it is complex, 
  the usual separation in spherical coordinates applies, for example:
 \begin{equation}
 F_A(E ; {\bf R})  = \sum_{J,M_J} \chi_{A,J} (k, R) Y_{JM_J} (\hat{\bf R}) \quad .
 \label{sph}
 \end{equation}
 We define the elastic-scattering wavenumber, $k_{A,J}(R)$,  for the incoming channel $A$ 
 with angular momentum $J$, as follows:
 \begin{equation}
 k_{A,J}^2(R) = k^2 -2\mu  V_A(R) -J(J+1)/R^2  \quad , 
 \end{equation}
 Then, without fear of ambiguity, we define the collision wavenumber:
 \begin{equation}
 k = \lim_{R \rightarrow \infty}  k_{A,J} \qquad .
 \label{wavk}
 \end{equation}
 Then the corresponding radial functions,  $f_{A,J}(k, R)=kR \chi_{A,J} (k, R)$, will be the solutions of the equations:
 \begin{equation}
\left[ \frac{d^2}{dR^2}+k_{A,J}^2(R)   \right] f_{A,J}(k, R) = 0 \quad .
\label{eqn9}
 \end{equation}
  normalized asymptotically ($R \rightarrow \infty$) according to,
 \begin{equation}
 f_{A,J}(k,R) \sim  \sqrt {\frac{2 \mu}{\pi k}}\sin  \left(  k R - \frac{1}{2}J\pi + \delta_J\right)
 \label{normc} 
 \end{equation}
and $\delta_J$ is the elastic  phase shift.  When the optical potential is used 
 the radial equations for the functions in (\ref{sph}) are the same:
 \begin{equation}
 \left[ {d^2 \over dR^2} +\kappa^2_{A,J}(R)   \right] \chi_{A,J} (k,R) =0 \quad ,
 \label{chiq}
 \end{equation}
 apart from the modification for the  complex wavenumber:
 \begin{equation}
\kappa^2_{A,J}(R) = k_{A,J}^2(R) -i\mu \Gamma(R) \quad . 
 \end{equation}
Since the imaginary term is short-ranged, then  $ \lim_{R \rightarrow \infty} \kappa_{A,J}(R) =k$ 
and the normalisation conventions for the radial wavefunctions  (\ref{normc}) are the same.  However, 
   $\chi_{A,J}(k,R)$ have  complex phase shifts  \cite{mott65} and thus the  probability flux is  attenuated. 
   
  Naturally  the vacuum emission represented by the width $\Gamma (R)$ is much smaller in magnitude 
  compared with the real potential $V_A(R)$ and thus we can solve (\ref{chiq})  by perturbation theory. 
 In the distorted-wave approximation  the imaginary part of the phaseshift
 ( $\mu_J = {\rm Im} \ \delta_J $)    is given by
 \begin{equation}
 \mu^{DW}_J(k) = \frac{\pi}{2} \int_{0}^{+\infty} | f_{A,J}(k, R)| ^ 2 \Gamma (R) dR
 \label{dwa}
 \end{equation}
 
We solve the problem directly  integrating (\ref{chiq}) using the Numerov method \cite{alli70,alli72,john77} 
and this is labelled the {\em quantal approximation} to distinguish it from the distorted-wave calculation 
and the semi-classical approximation discussed above. 

 The cross section for total collision-induced radiative decay from the entrance channel, the sum of the cross sections
for processes (\ref{proc1}) and (\ref{proc2}) can be obtained within the optical potential approximation.
 The cross section for collision-induced radiative decay can then be written as,
\begin{equation}
\sigma (E) = \frac{g\pi}{k^2} \sum_{J=0}^{\infty} (2J + 1){\left[ 1 -  e^{-4 \mu_J} \right]} \quad .
\label{quantal}
\end{equation}
where $k$ is given by (\ref{wavk}), and $g$ is the spin (statistical) weight.  Since the loss channel  is via the   $X ^1\Sigma^+$ state, 
and as the Yb ion and Rb atom combine to produce singlets, then only  ion-atom collisions with singlet symmetry 
have a dipole-allowed spontaneous emission. So in this case, the statistical weight is, $g=1/4$.

At higher energy,  a semi-classical approximation is invoked to calculate the cross sections for radiative
decay.  The summation over the angular momentum in equation \ref{quantal} can be replaced by an integral over the 
impact parameter, $b$, according to $ k b \approx J$. 
The JWKB approximation can then be used to obtain the wave function, 
\begin{equation}
f_{A,J}(k, R) \approx  \sqrt {\frac{2 \mu}{\pi k_{A,J} (R)}}\sin  \left( \int_{R_c}^R k_{A,J}(R') dR' +\textstyle{ 1 \over 4} \pi \right)  \quad .
\end{equation}
This  simplifies the calculation of the phase-shift , equation (\ref{dwa})
  \cite{zyge88,alli66,bates51}  since the rapidly varying integrand gives us 
  (in the classically allowed region):
$ f^2_{A,J}(k, R) \approx   \mu/(\pi k_{A,J}(R))  $. Then 
using (\ref{dwa}) we get the semi-classical approximation (\ref{semic}).

 
 The thermally averaged rate coefficient $\alpha (T) =\langle v \sigma \rangle $, as a function of temperature $T$, is obtained by averaging 
  over the  Maxwell-Boltzmann distribution. That is, 
 \begin{equation}
 \alpha (T) = \left( { 8 \over \mu \pi k_{B}^{3} T^{3}} \right)^{1/2}  \int_{0}^{\infty} E ~ \sigma (E) e^{- E/(k_B T)}  d E.
\label{rate}
 \end{equation}
In the early work of  Bates  \cite{bates51} an efficient and convenient procedure to evaluate the rate coefficient was outlined. 
 In the present calculations for cross sections we start from 10$^{-12}$  $\mu$eV and extend these to higher energies,
  by invoking  a semi-classical approximation above about 10$^{-2}$ eV up to 10$^4$ eV for the transition of interest.
 \begin{figure}
\includegraphics[width=0.525\textwidth]{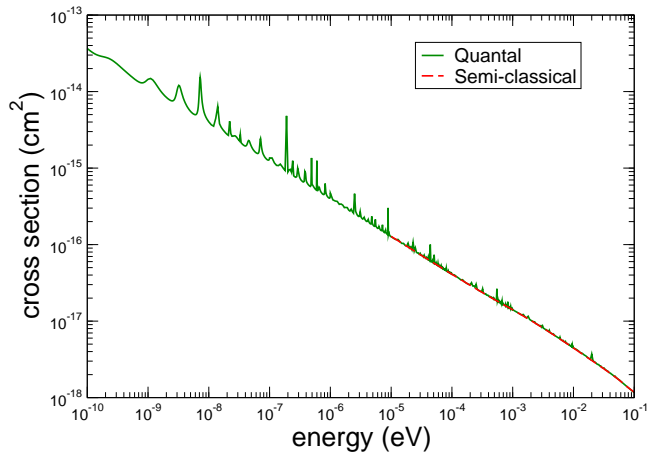}
\caption{
Cross sections for the total collision-induced radiative decay  at low energies (\ref{proc1}) for $^{174}{\rm Yb}^{+} +{\rm Rb}$.
The quantal optical potential calculations (\ref{quantal}) are compared with  the semi-classical approximation. 
In the figure we present the spinless cross sections, that is $g=1$.
The background of the quantal result (\ref{quantal}) follows the semi-classical  
approximation (\ref{semic}) at collision energies above $\mu$eV  and has the asymptotic $E^{-1/2}$ behaviour. }
\label{fig5}
\end{figure}
  \section{Results} 
 \subsection{Electronic states}
  Since the MRCI calculations do not explicitly include relativistic effects, although this is not
important for the entrance collision channel or the lower Yb ($^1$S) + Rb$^+$ ($^1$S) asymptote as all the
molecular states formed are of $\Sigma^+$ symmetry. This is borne out by the calculated energy of the
asymptotic energies of the a$^3\Sigma^+$ and A$^1\Sigma^+$ states \cite{lamb12}.
The asymptotes for the higher $^3\Pi$ and $^3\Sigma^+$ states 
correlate  to the Yb (6s6p~$^3$P$^o$) + Rb$^+$ (4p$^6$~$^1$S) atomic products. The multiplet 
and its associated fine-structure splitting in the triplet (Yb: $^3$P$^o_{0,1,2}$)  is considerable: $\sim$0.3 eV. 
Only a fully relativistic treatment can 
accurately account for the spin-orbit interaction. In a magnetic trap of course the Zeeman splitting and hyperfine structure complicates 
matters further. Nonetheless, in our first analysis of this novel system, we can confidently say 
that a curve crossing will take place between the A$^1\Sigma^+$ and b$^3\Pi$ states
though at an energy above the Yb$^+$ ($^2$S) + Rb ($^2$S) asymptote. Such a crossing will facilitate a charge exchange
reaction as observed in experiment at mK temperatures~\cite{ zipk10b,zipk10a}. In our previous work on this complex \cite{lamb12} 
we have estimated the molecular constants for the four states that support bound rovibrational states. 

\begin{figure}
\includegraphics[width=0.525\textwidth]{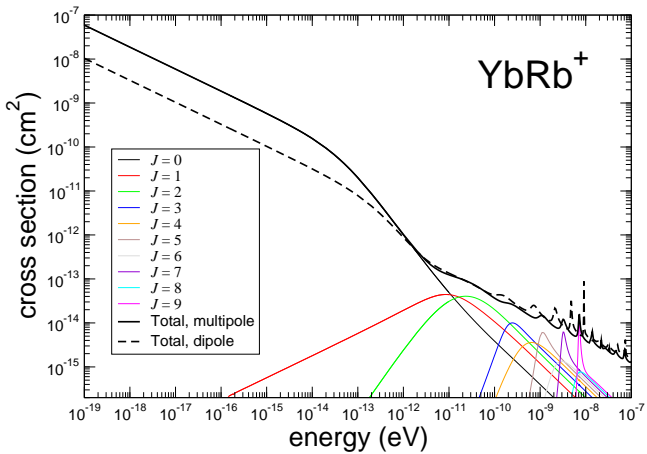}
\caption{Partial cross sections for radiative charge transfer and radiative association for
	       the sum of reactions (1) and (2) as a function of relative collision energy
	       for the $\rm^{174}Yb$ isotope. The contribution of each partial wave is shown, and 
	       illustrates the sharp potential  resonances which are tuned by the centrifugal barrier.
	       The solid line and dashed line show the effect of different long-range interactions. 
	       Referring to (\ref{mult}),  the solid line is the full multipole expansion, while the dashed line only includes 
	       the dipole term, that is $C_6=C_8=0$. Again, in this case we present the spinless partial 
	       cross section, equation (\ref{quantal}),  with $g=1$.
	        }
\label{fig6}
\end{figure}
 \begin{figure}
\includegraphics[width=0.525\textwidth]{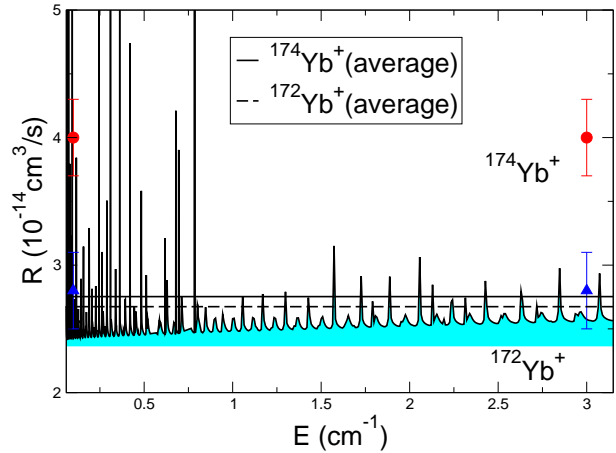}
\caption{The energy-dependent effective rate $R (E)$ (cm$^3$/s) = $v \sigma$, defined in (\ref{crate}), 
		compared with the available 
		experimental data on the two different isotopes of the Yb$^+$ ion,
		(solid blue triangle $\rm^{172}Yb^+$, solid red circle $\rm^{174}Yb^+$).  
		Mean theoretical value are shown from the present optical potential calculations 
		for the two different isotopes of the Yb$^+$ on. The solid black line is the average 
		through the resonances for the  case of the $\rm^{174}Yb^+$ isotope,
		the dashed line that for the  $\rm^{172}Yb^+$ isotope.  Although theory shows
		suitable agreement with the experimental measurements, 
		the strong isotope sensitivity observed in the experiment is 
		not evident in the present calculations.
		}
\label{fig7}
\end{figure}

\subsection{Cross sections and collision rates} 
 Cross sections were determined using the quantal optical potential approximation, 
 for collision energies ranging from 10$^{-12}~\mu$eV up to 10 eV.  
At  higher collision energies a semi-classical approximation (Eq. 12) was invoked for energies 
up to 10 keV in order to determine the cross sections.  

In order to compare with experiment the statistical weight for the singlet, must be taken 
into account, this means  taking $g=1/4$ in (\ref{quantal}).  The cross section results  presented in 
Fig. \ref{fig5}  and Fig. \ref{fig6} for  all the radiative decay process are the spinless results, $g=1$.  
In Fig. \ref{fig5} we show the optical potential results 
as they are mapped on to those obtained from the semi-classical approximation.  Fig. \ref{fig6} illustrates 
 the low partial-wave contributions to the total cross section in the optical potential approximation.  
 The solid line and dashed line show the effect of the long-range interactions. 
 The solid line is the full multipole expansion, while the dashed line only includes the quadrupole of polarization. 
 At the higher energies, the corrections to leading-order  polarization do not affect the positions of the resonances. 
 This confirms that the resonance  effect as short-range features, well described as potential scattering.  
 At the very lowest energies however, as the wavelengths become extremely long and the 
 centrifugal barrier much more significant, then, as is well-known, the long-range features of the potential take over.
From these results one clearly sees that at collision energies below 10$^{-6}~\mu$eV the cross 
section is totally dominated by $s$-wave scattering. However, owing to the sensitivity 
of the scattering to the potential, the estimation of the effective range parameters, including 
the complex scattering length, would be of great interest. 
So primarily, at intermediate energies the process is dominated by the short-range 
classical turning point. This may go some way to explaining why our dynamic 
results are in such good agreement with the more complex calculations 
of  \cite{sayf13} although the agreement is surprising.  
Note, at higher energies non-adiabatic effects will naturally become more important,
however, in this energy (temperature) range, the role of non-adiabatic 
coupling (radial and rotational) turns out  to have very little importance indeed. 
Their influence is negligible for the radiative capture process, as has been
shown recently in detailed studies by Sayfutyarova et al.  \cite{sayf13}.

Finally, we  consider whether thermal effects might be taken into account, 
to confirm the discrepancy between theory and experiment.
In the experiments by K\"ohl et al. \cite{zipk10a,zipk10b} the kinetic
energy of a single Yb$^+$ ion immersed in an ultracold Rb
ensemble was varied by adding excess micromotion energy
after displacement of an ion from the centre of a trap. The
binary-collision ion-loss rate coefficient determined in this
way does not correspond to a conventional thermally-averaged rate
constant for the Maxwell collision energy distribution at a certain temperature which assumes 
thermal equilibrium.
 The relationship $v \sigma$ is therefore used to designate an effective energy-dependent 
 rate coefficient  \cite{sayf13}, where  $R (E)$  is given by, 
 \begin{equation}
 R(E) = \sqrt{2E/\mu}\;  {\sigma}^R_{A \rightarrow X} (E)
 \label{crate}
 \end{equation}
 We use this form to define a quasi-rate coefficient \cite{krems10} rather than one averaged over 
 a Maxwellian distribution defined in equation \ref{rate}.
 In figure Fig. \ref{fig7} we compare our calculations with experiment for this quasi-rate parameter $R$. 
The measured experimental value for the $^{174}{\rm Yb}^+$ isotope \cite{zipk10b} shown in figure Fig. \ref{fig7} 
indicate that the magnitude of $R(E)$ is  $ (4.0 \pm 0.3) \times 10^{-14}$ cm$^3$/s, where as the calculations of 
Sayfutyarova et al.  \cite{sayf13} for this isotope  give a value, 
after averaging the cross sections in the energy region 0.15 -- 3.25 cm$^{-1}$, 
through the  savannah of resonances,  a value 
of  $2.9\times 10^{-14} $ cm$^3$/s,  which is just outside the experimental error. 
Carrying out a similar procedure with our cross sections results (solid black line, Fig. \ref{fig7})
yields a mean value slightly higher, in better agreement with experiment, 
but relatively close in magnitude to previous work \cite{sayf13}. 
For the $^{174}{\rm Yb}^{+}$ isotope  we obtained a value of $R(E) \approx 2.76 \times 10^{-14} $cm$^3$/s. 
Similarly, for the isotope $^{172}{\rm Yb}^{+}$,  (after averaging through the resonances features, dashed black line, Fig. \ref{fig7})
we obtained a value $\approx 2.68 \times 10^{-14} $cm$^3$/s,  once again close to the results of Sayfutyarova et al.  \cite{sayf13}.
Experimental studies show there is a large isotope shift, as measurements indicate a value
of $R(E)$ for the $^{172}{\rm Yb}^+$  isotope  of $(2.8 \pm 0.3) \times 10^{-14}$ cm$^3$/s  \cite{zipk10b} with the ratio
$R_{174}/R_{172} \approx 1.4$. As found in  previous studies  \cite{sayf13} our theoretical predictions for this same ratio give a value 
of $\approx$ 1.03 indicating to the contrary.

Regarding the large isotope sensitivity, we do not observe as great a difference as reported in
experimental measurements.   Similar to the detailed calculations of Sayfutyarova et al.  \cite{sayf13} 
we find a dense forest of resonances, but no broad features 
that would lead to a strong isotope dependency as observed in the experiment. So, our conclusion 
is that this feature remains unexplained. It is possible to speculate that the experimental conditions, 
having the magnetic field present, create additional complications. For example 
the nuclear spin of the Yb ion. It is known from recent experiments that 
this has an important role in relaxation phenomena \cite{kohl13}. Further detailed theoretical studies 
and additional experiments would be essential in order to resolve this issue.

We note for YbCa$^{+}$ ultra-cold collisions \cite{zyge14} it is only at temperatures below a nano-Kelvin (10$^{-6}$ K)
that a large isotope effect is seen. Above these temperatures there is a very small isotope effect.  
In the present work on YbRb$^{+}$, the energy range is 0.15 - 3.25 cm$^{-1}$ (0.215 -  4.676 Kelvin), 
so a similar small isotope effect is seen as in YbCa$^{+}$ \cite{zyge14}.

\section{Conclusions}
We have investigated the quantum nature of ultracold ion-atom collisions and calculated 
the relevant experimental processes - cross sections and rate coefficients for the different isotopes of Yb$^+$ ion.
These calculations are important in the design and interpretation of the new generation 
of experiments  involving ultracold ion-atom systems.
Potential energy curves and transition dipole moments obtained from the MOLPRO suite of codes 
for low-lying molecular states of the diatomic molecular ionic system 
containing a ytterbium ion and a rubidium atom, with relevance 
to ultra-cold chemistry were used in our dynamical calculations. 
Cross sections as a function of energy, for the radiative decay, charge transfer 
and association processes involving Yb ions and Rb 
atoms are determined using an optical potential method. 
The multi-reference configuration interaction (MRCI) 
approach is used to determine turning points, crossing points, potential minima 
and spectroscopic molecular constants 
 for the  lowest five molecular states.  The long-range parameters, including  
the dispersion coefficients estimated from our {\it ab initio} data  
were used in our dynamical investigations.
Quasi-energy dependent rate coefficients are determined from our cross section for the radiative decay processes in ultracold 
collisions of  a ytterbium ion and a rubidium atom based on our  {\it ab initio}  data  and compared 
with the available experimental measurements and previous theoretical work  \cite{sayf13}.  The agreement 
is surprisingly good for the molecular electronic structure that gives rise to the complex optical potential.
The smooth nature of the optical potential and validity of the semi-classical approximation indicates that 
one can accurately estimate the cross section with an elementary  quadrature. The more complex quantal treatment,
while exhibiting the expected potential resonances, does not  give rise to a strong isotope effect, at least 
in the energy range we investigated. The estimates of the energy dependent collision rate are in suitable agreement 
with experiment  \cite{zipk10a,zipk10b} and with previous theoretical studies \cite{sayf13}. We find no broad resonances 
features that might underly a strong isotope effect. In conclusion, we find, in agreement with 
previous theoretical work  \cite{sayf13}  that the isotope anomaly observed in experimental  studies remains unexplained. 
%
%
%
%
\begin{acknowledgments}
HDLL is grateful to the Department of Employment and Learning (DEL, Northern Ireland) for the provision of a
postgraduate studentship.  B MMcL thank Queen's University Belfast for the award of a Visiting Research Fellowship  
and  for support by the US National Science Foundation under the visitors program through a grant to ITAMP
at the Harvard-Smithsonian Center for Astrophysics.   Grants of computational time at the National Energy 
Research Scientific Computing Center in Oakland, CA, USA and at the High Performance Computing Center Stuttgart
(HLRS) of the University of Stuttgart, Stuttgart, Germany are gratefully acknowledged.
\end{acknowledgments}
%
%
%
%
\bibliographystyle{apsrev4-1}
\bibliography{ybrb}
\section*{Appendices}
\renewcommand{\thetable}{\Alph{subsection}.\arabic{table}}
\setcounter{table}{0}
\subsection{YbRb$^+$ MRCI potentials and dipole moments}
\begin{table*}
\caption{\label{tab1} Energies for the lowest five states of the YbRb$^+$ cation and the magnitude of the transition
				dipole moments $|\mu|$ (all values in  atomic units) between the $^{1,3}\Sigma^{+}$ states. Calculations were performed 
				at the multi-reference-configuration-interaction (MRCI)  level as a 
				function of internuclear separation R (a$_0$) with the MOLPRO suite of codes \cite{Werner2010}, see text for details.
				The  potential energy curves (PEC's) are calculated using an effective core 
				potentials  (ECP) to replace the non-valence electrons (ECP68MDF for Yb, ECP36SDF for Rb) and an AV6Z basis
				for the outer electrons.}
\begin{ruledtabular}
\begin{tabular}{cccccccc}
\\
R(a$_0$) &X$^1\Sigma^{+}$  & A$^1\Sigma^{+}$	&a$^3\Sigma^{+}$&2$^3\Sigma^{+}$&b$^3\Pi$
&$| \mu_{X ^1\Sigma^{+}  \leftarrow A ^1\Sigma^{+}} |$	
& $| \mu_{a ^3\Sigma^{+} \leftarrow 2 ^3\Sigma^{+}} |$ \\
\\
\hline
\\
     3.0    &-0.32382167   & -0.18056641  &-0.22685070   &-0.13830772   &-0.23369659   &2.77538976E+00    &3.14573472\\ 
     3.5    &-0.40254912   &-0.26228838   &-0.30820703   &-0.22197268   &-0.31847629   &2.78930438E+00    &3.18660134\\
     4.0    &-0.46032058   &-0.32219208   &-0.36780910   &-0.28305587   &-0.38021812   &2.79840138E+00    &3.22631386\\
     4.5    &-0.50698122   &-0.37077020   &-0.41545887   &-0.33254144   &-0.42766881   &2.82154025E+00    &3.27548581\\
     5.0    &-0.54556668   &-0.41162076   &-0.45527239   &-0.37539631   &-0.46478422   &2.85647641E+00    &3.31650415\\
     5.5    &-0.57635600   &-0.44539034   &-0.48851320   &-0.41236776   &-0.49339932   &2.88180744E+00    &3.33737913\\
     6.0    &-0.59936584   &-0.47200281   &-0.51535252   &-0.44216198   &-0.51463657   &2.89339159E+00    &3.35410838\\
     6.5    &-0.61530769   &-0.49190959   &-0.53595262   &-0.46446248   &-0.52956862   &2.89862641E+00    &3.38211334\\
     7.0    &-0.62556300   &-0.50637097   &-0.55101842   &-0.48054432   &-0.53947088   &2.90026933E+00    &3.43126163\\
     7.5    &-0.63170310   &-0.51696730   &-0.56165122   &-0.49223745   &-0.54569486   &2.89332163E+00    &3.50223695\\
     8.0    &-0.63506849   &-0.52507544   &-0.56893750   &-0.50108003   &-0.54941361   &2.87174916E+00    &3.59210829\\
     8.5    &-0.63663599   &-0.53164742   &-0.57374404   &-0.50810491   &-0.55150557   &2.83406169E+00    &3.69858005\\
     9.0    &-0.63707106   &-0.53721379   &-0.57671392   &-0.51394276   &-0.55257977   &2.78292776E+00    &3.82000620\\
     9.5    &-0.63682035   &-0.54200285   &-0.57832056   &-0.51896732   &-0.55304467   &2.72098159E+00    &3.95539621\\
    10.0   &-0.63618486   &-0.54608639   &-0.57891704   &-0.52339749   &-0.55316881   &2.64796384E+00   &4.10437100\\
    11.0   &-0.63449906   &-0.55222523   &-0.57808634   &-0.53091970   &-0.55301220   &2.45740465E+00   &4.44339547\\
    12.0   &-0.63290464   &-0.55591942   &-0.57570186   &-0.53698444   &-0.55279144   &2.19636550E+00   &4.83663826\\
    13.0   &-0.63168317   &-0.55771850   &-0.57265590   &-0.54178475   &-0.55267624   &1.87938443E+00   &5.28052502\\
    14.0   &-0.63083565   &-0.55826383   &-0.56949338   &-0.54548429   &-0.55265948   &1.54260849E+00   &5.76703935\\
    15.0   &-0.63027159   &-0.55810404   &-0.56653201   &-0.54826080   &-0.55270028   &1.22153841E+00   &6.28713063\\
    16.0   &-0.62989719   &-0.55762152   &-0.56392949   &-0.55029189   &-0.55276659   &9.38646894E-01    &6.83305805\\
    17.0   &-0.62964301   &-0.55704497   &-0.56173872   &-0.55173950   &-0.55284007   &7.03010256E-01    &7.39830456\\
    18.0   &-0.62946444   &-0.55649325   &-0.55995322   &-0.55274269   &-0.55291181   &5.14825073E-01    &7.97609013\\
    19.0   &-0.62933454   &-0.55601681   &-0.55853607   &-0.55341611   &-0.55297811   &3.69225621E-01    &8.55803409\\
    20.0   &-0.62923721   &-0.55562762   &-0.55743586   &-0.55385114   &-0.55303782   &2.60513473E-01    &9.13946304\\
    21.0   &-0.62916259   &-0.55531895   &-0.55659649   &-0.55411837   &-0.55309091   &1.81066506E-01    &9.71503237\\
    22.0   &-0.62910436   &-0.55507720   &-0.55596395   &-0.55427057   &-0.55313784   &1.24246627E-01    &10.28412960\\
    23.0   &-0.62905827   &-0.55488807   &-0.55549067   &-0.55434598   &-0.55317921   &8.42909856E-02    &10.85084727\\
    24.0   &-0.62902137   &-0.55473920   &-0.55513739   &-0.55437148   &-0.55321565   &5.66083918E-02    &11.42134629\\
    25.0   &-0.62899152   &-0.55462086   &-0.55487342   &-0.55436535   &-0.55324775   &3.76692214E-02    &11.98876546\\
    26.0   &-0.62896717   &-0.55452571   &-0.55467596   &-0.55433935   &-0.55327608   &2.48540266E-02    &12.46999337\\
    27.0   &-0.62894713   &-0.55444834   &-0.55452916   &-0.55430013   &-0.55330112   &1.62676499E-02    &12.51192229\\
    28.0   &-0.62893053   &-0.55438476   &-0.55442250   &-0.55425082   &-0.55332327   &1.05572203E-02    &11.29536440\\
    29.0   &-0.62891668   &-0.55433202   &-0.55434731   &-0.55419427   &-0.55334293   &6.80540314E-03    &8.50476276\\
    30.0   &-0.62890504   &-0.55428792   &-0.55429362   &-0.55413588   &-0.55336041   &4.35528758E-03    &5.49589293\\
    31.0   &-0.62889521   &-0.55425076   &-0.55425284   &-0.55408066   &-0.55337599   &2.76768243E-03    &3.32888059\\
    32.0   &-0.62888686   &-0.55421927   &-0.55422002   &-0.55403074   &-0.55338993   &1.74670931E-03    &1.98772267\\
    33.0   &-0.62887973   &-0.55419241   &-0.55419268   &-0.55398638   &-0.55340241   &1.09489525E-03    &1.18859254\\
    34.0   &-0.62887360   &-0.55416939   &-0.55416949   &-0.55394711   &-0.55341363   &6.81718585E-04    &0.71360777\\
    35.0   &-0.62886832   &-0.55414956   &-0.55414959   &-0.55391234   &-0.55342373   &4.21603382E-04    &0.42971578\\
    36.0   &-0.62886375   &-0.55413240   &-0.55413241   &-0.55388147   &-0.55343285   &2.58956635E-04    &0.25907031\\
    37.0   &-0.62885977   &-0.55411748   &-0.55411749   &-0.55385398   &-0.55344111   &1.57915736E-04    &0.15609902\\
    38.0   &-0.62885630   &-0.55410448   &-0.55410448   &-0.55382943   &-0.55344860   &9.56016471E-05    &0.09385741\\
    39.0   &-0.62885325   &-0.55409310   &-0.55409310   &-0.55380743   &-0.55345541   &5.74329870E-05    &0.05623899\\
    40.0   &-0.62885058   &-0.55408310   &-0.55408310   &-0.55378765   &-0.55346161   &3.42271192E-05    &0.03354626\\
    42.0   &-0.62884612   &-0.55406649   &-0.55406649   &-0.55375374   &-0.55347244   &1.18506542E-05    &0.01173501\\
    44.0   &-0.62884262   &-0.55405342   &-0.55405342   &-0.55372591   &-0.55348154   &3.95589429E-06    &0.00399752\\
    46.0   &-0.62883983   &-0.55404304   &-0.55404304   &-0.55370287   &-0.55348922   &1.27549229E-06    &0.00132327\\
    48.0   &-0.62883758   &-0.55403469   &-0.55403469   &-0.55368363   &-0.55349576   &3.93135974E-07    &0.00042512\\
    50.0   &-0.62883576   &-0.55402793   &-0.55402793   &-0.55366744   &-0.55350136   &1.16661070E-07    &0.00013179\\
                \\
\end{tabular}
\end{ruledtabular}
\end{table*}

\end{document}